\documentclass[fleqn,10pt]{wlscirep}
\usepackage[utf8]{inputenc}
\usepackage[T1]{fontenc}
\usepackage{graphicx}
\usepackage{physics}
\usepackage{bm}
\title{Subcarrier wave continuous variable quantum key distribution with discrete modulation: mathematical model and finite-key analysis}

\author[1,2,*]{E.O. Samsonov}
\author[1]{R.K. Goncharov}
\author[1,2]{A.A. Gaidash}
\author[1,2]{A.V. Kozubov}
\author[1,2]{V.I. Egorov}
\author[1]{A.V. Gleim}

\affil[1]{ITMO University, Kronverkskiy, 49, Saint Petersburg, 197101, Russia}

\affil[2]{Quanttelecom LLC., Saint Petersburg, 199178 6 Line 59, Russia}

\affil[*]{eosamsonov@itmo.ru}

\begin{abstract}
In this paper we report a continuous-variable quantum key distribution protocol using multimode coherent states generated on subcarrier frequencies of the optical spectrum. To detect the quadrature components of bosonic field we propose a coherent detection scheme where power from a carrier wave is used as a local oscillator. We compose a mathematical model of the proposed scheme and perform its security analysis in the finite-size regime using fully quantum asymptotic equipartition property technique. We calculate a lower bound on the secret key rate for the system under the assumption that the quantum channel noise is negligible compared to detector dark counts, and an eavesdropper is restricted to collective attacks. Our calculation shows that the current realistic system implementation would allow distributing secret keys over channels with losses up to 9 dB.
\end{abstract}
\begin{document}

\flushbottom
\maketitle

\thispagestyle{empty}

\section*{Introduction}

Quantum key distribution (QKD) is a method of sharing symmetric cryptographic keys between two parties that is based on encoding information in the states of quantum objects and subsequent distillation of the key through a classic communication channel. The first quantum cryptography protocols exploited the quantum system with  degrees of freedoms \cite{Bennett2014, Bennett1992, BennetC.H.1992}. A numerous amount of different techniques for security proofs for discrete variable QKD systems has already been presented \cite{Tamaki2003,Christandl2004,Renner2005,christandl2009postselection,QAEP,Renner2008,kraus2005lower,lo1999unconditional,shor2000simple,lo2005decoy}. Experimental implementations of this family of QKD protocols rely on single-photon detectors for quantum state measurements. 

In turn, continuous-variable QKD (\mbox{CV-QKD}), which was proposed later, relies on methods of coherent detection, homodyne or heterodyne, for gaining information about the quantum states. In other words, single-photon detection is replaced by conventional optical communication methods. However, security proofs for CV-QKD protocols currently remain less advanced \cite{Pirandola2019, Scarani2009}. 

There are two types of CV protocols that differ by signal modulation method: Gaussian \cite{Grosshans2002, Grosshans2003}, where the complex amplitudes of coherent states are selected randomly from a normal distribution, and discrete modulation (DM) \cite{Hirano2003, Leverrier2011, Heid, Bradler2018, Papanastasiou2018} with weak coherent phase-coded states. Other \mbox{CV-QKD} protocols are based on two-mode squeezed vacuum states transmission and measurement via homodyne or heterodyne detection \cite{Cerf2001}. Security proofs for Gaussian CV-QKD protocols remain the most developed: they were presented against general attacks in the finite key regime using several different approaches \cite{Diamanti2015}. Security analysis for \mbox{CV-QKD} protocol with two-mode squeezed vacuum states was also performed \cite{Guang-Qiang2008, Madsen2012}. Discrete-variable \mbox{CV-QKD} protocols possess several important advantages; among those are relative implementation simplicity and a possibility to minimize the number of parameters that need to be monitored. Nevertheless, security proofs for discrete-modulation \mbox{CV-QKD} systems require special consideration. In the asymptotic limit, its security has been proven against collective attacks \cite{Ghorai2019}. Recently it was shown that security proof for \mbox{CV-QKD} with discrete modulation against general attacks is possible \cite{Ghorai2019}. 

Here we propose an implementation of CV-QKD protocol based on subcarrier wave (SCW) technique \cite{Merolla1999,Mora2012,Gleim2016,gleim2017sideband,glejm2014quantum,mel2018using,gaidash2019methods,gaidash2019countermeasures,Kozubov2017}. A defining property of subcarrier wave DV-QKD is the method for quantum state encoding. In it, a strong monochromatic wave emitted by a laser is modulated in an electro-optical phase modulator to produce weak sidebands, whose phase with respect to the strong (carrier) wave encodes quantum information (for more details, see \cite{Gleim2016}). Like in any other DV-QKD systems, in SCW QKD the weak radiation component is detected by a single photon counter, and the measured observable has a discrete spectrum. In SCW CV-QKD protocol described in this work, Alice prepares coherent multimode states, which can be defined as quadratures of the bosonic field, while Bob performs coherent detection to establish correlations with Alice. 

Here we propose a new scheme for detecting quadrature components of the bosonic field, the main advantage of which is using the carrier wave (an essential part of SCW methodology) as a local oscillator. In practice, it solves the well-known problem of transmitting the local oscillator through the quantum channel (or its generation on receiver’s side). This is a novel approach that has not been discussed in previous works dedicated to studying multimode CV QKD.\cite{fang2014,gyongyosi2014a,wang2019}. 

From telecommunication point of view, SCW approach possesses several additional advantages. Firstly, it is intrinsically robust against external conditions affecting the fiber and is ready to function in conventional telecom infrastructure. Secondly, it demonstrates unmatched spectral efficiency in the quantum channel, allowing for distributing several keys on separate closely-packed sidebands around a single optical carrier \cite{Mora2012}. Thirdly, recent experiments \cite{kynev2017free} have shown that preservation of SCW quantum signal parameters in respect to the carrier allows transmitting phase-encoded quantum signals through the air providing invariance to telescope rotation that remains an important obstacle in traditional polarization-based free-space quantum communication, making the same QKD kit suitable for fiber and free-space QKD networks. Security proof of SCW QKD protocol with discrete variables against collective beam-splitting attack was proposed in \cite{Kozubov2017}, and more recently general finite-key security proof was presented in \cite{Kozubov2019}.

A major difference of SCW approach from the previous CV-QKD protocols is using multimode coherent states generated on subcarrier frequencies. It therefore requires special consideration of security proof technique for the CV-QKD protocol. The most advanced security descriptions for typical CV-QKD protocols with Gaussian and discrete modulation assume that the quantum channel has losses and imposes Gaussian noise on the observed quadrature distributions. For CV-QKD this usually requires estimating a covariance matrix of the bipartite state shared by Alice and Bob \cite{Diamanti2015}. In Gaussian modulation protocols the variances and covariances directly measured by Alice and Bob give a covariance matrix. In case of DM protocols it is harder to obtain, but in \cite{Ghorai2019} a major step towards the full security proof of DM CV-QKD has been presented. The lower bound against collective attacks is calculated by solving a semidefinite program that computes the covariance matrix of the state shared by Alice and Bob in the entanglement-based version of the protocol. Our aim in this work is to demonstrate universality of CV-QKD protocol based on SCW technique. Hence we build a mathematical model of CV-QKD protocol based on SCW method and show the possibility of performing security proof analysis in case of multimode coherent states. Unconditional security proof is out of scope of this paper and will be a subject for a separate study. Here we perform finite-key security analysis using fully quantum asymptotic equipartition property technique \cite{QAEP} and calculate the lower bound on secret key rate  under the assumption that detector dark counts remain a dominant contribution to the total noise level \cite{Heid}. The key rates are obtained for direct reconciliation scheme with post-selection in case of collective attacks.

\section*{\label{sec:part2}Results}

\subsection*{\label{sec:part21}Subcarrier wave CV-QKD setup}

\begin{figure}[ht!]
\includegraphics[width=\textwidth]{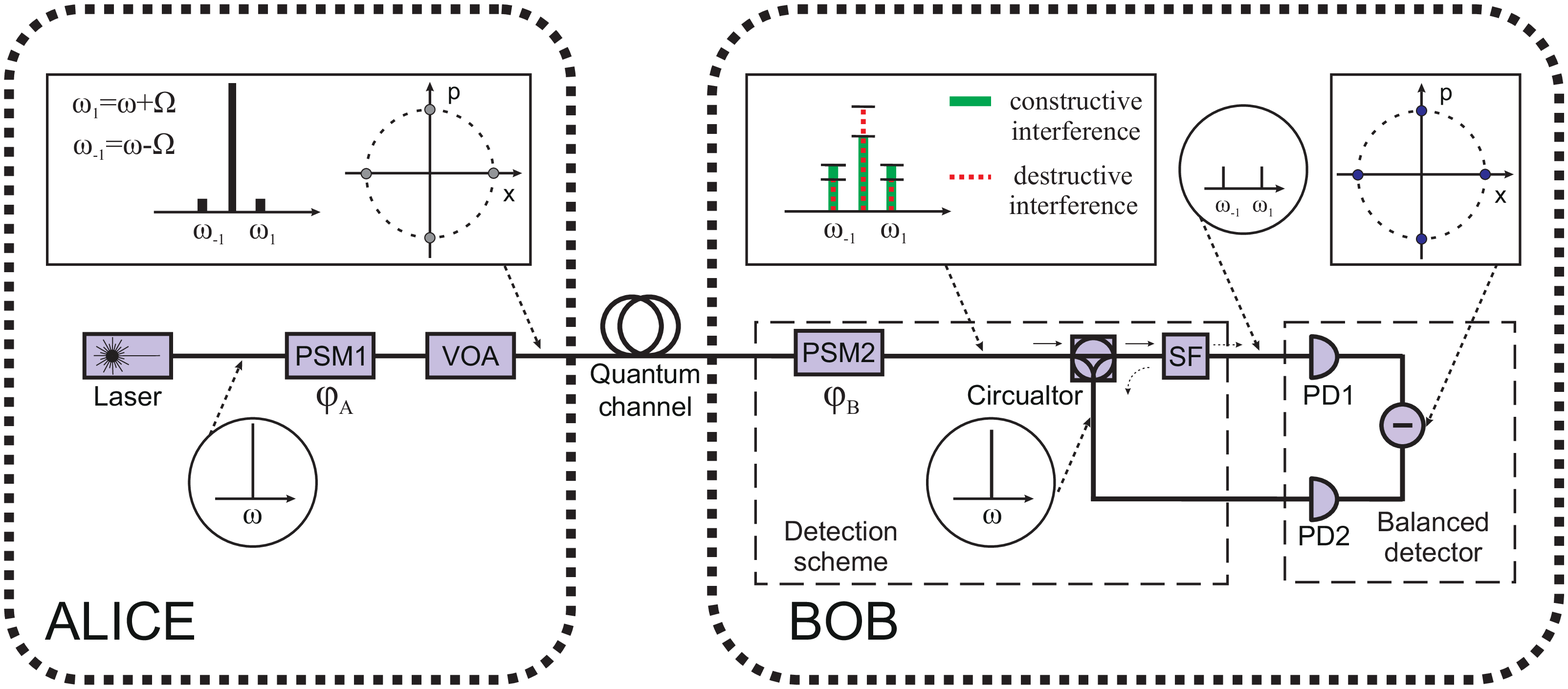} \caption{Principal scheme of SCW CV-QKD setup. PSM is an electro-optical phase modulator; VOA is a variable optical attenuator; SF is a spectral filter that cuts off the carrier; PD is a photodiode. Diagrams in circles show the absolute value of signal spectrum taking into account only the first-order subcarriers. Diagrams in squares illustrate the absolute value of signal spectrum and comparison of spectra for various phase shifts; different coherent states are shown on phase plane.} 
\label{fig1}
\end{figure}

In SCW method the signal photons are not emitted directly by a laser source but are generated on subcarrier frequencies, or sidebands, in course of phase modulation of an intense optical carrier. Laser source emits coherent light with frequency $\omega$. Alice modulates this beam in a traveling wave electro-optical phase modulator with the microwave field with frequency $\Omega$ and phase $\varphi_{A}$ \cite{Yariv1984}. As a result, pairs of sidebands are formed at frequencies $\omega_{k}=\omega+k\Omega$, where integer $k$ runs between the limits: $-S\leq k \leq S$. Modulation index at Alice side is chosen so that the total number of photons in the sidebands is less than unity (according to the QKD protocol). 
In the proposed SCW CV-QKD setup shown in Fig. \ref{fig1} Alice sends weak coherent states along with the carrier through a quantum channel. Alice prepares her states using quadrature phase-shift modulation by choosing from a finite set of states $\varphi_A\in\{0,\:\pi/2,\:\pi,\:3\pi/2\}$. Receiver (Bob) applies much higher modulation index than Alice on his modulator and randomly selects $x$ or $p$ measurement introducing phase shift $\varphi_B\in\{0,\:\pi/2\}$, respectively, in each transmission window $T$.
Here we consider CV-QKD protocol with discrete modulation, so we formally leave Alice's block the same as in initial DV-QKD system \cite{Gleim2016}, but completely change the detection scheme.

\begin{figure}[ht!]
\includegraphics[width=\textwidth]{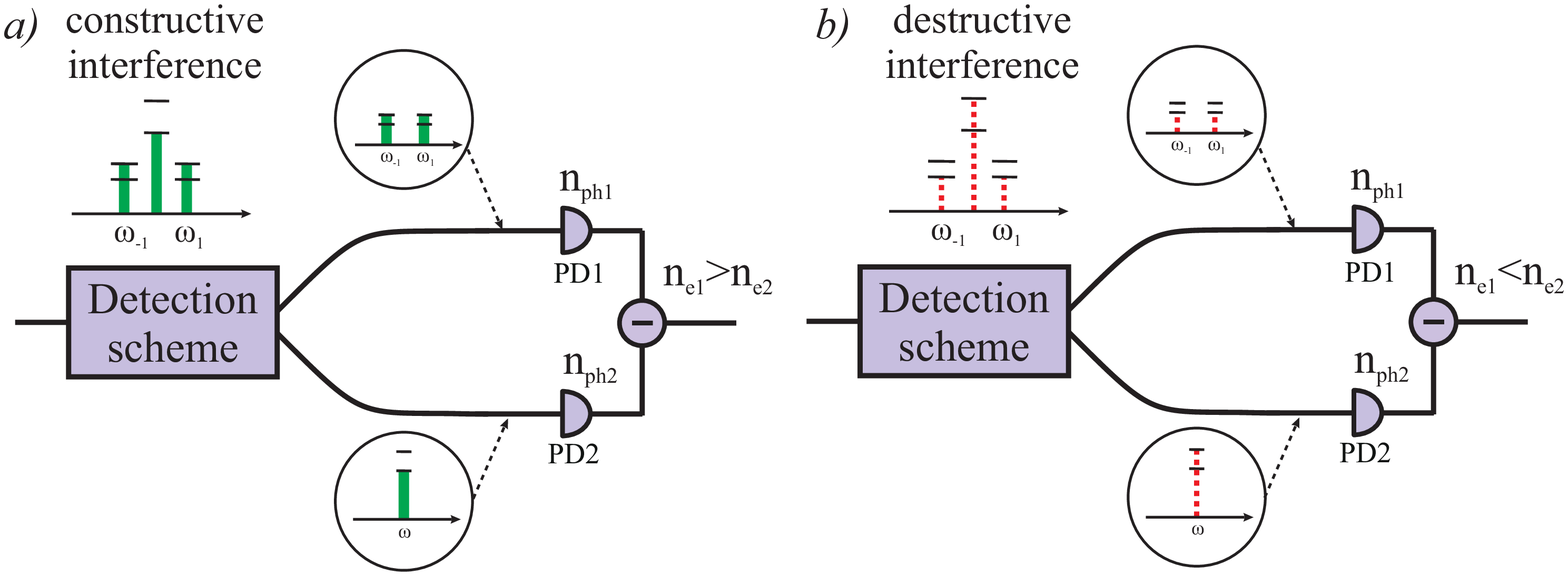} \caption{SCW coherent detection scheme operation. The charts show energy distribution between the carrier and the subcarriers in case of constructive (a) and destructive (b) interference. Subcarrier signal power becomes  higher or lower than the carrier power, respectively. Horizontal dashes added for illustrative purposes.}
\label{fig1.1}
\end{figure}

Fig. \ref{fig1.1} describes the operation of proposed coherent detection scheme in detail. We avoid mentioning the words "homodyne" and "heterodyne" purposely because this paper does not consider a classical scheme, but its analog, corresponding to the more general definition of "coherent detection". By definition, homodyne detection is characterized by interference of a weak signal with a powerful local oscillator on a $50/50$ beam splitter. After interference, the number of photons at the detectors $n_1$ and $n_2$ depends on phase difference $\Delta=\phi_{A}-\phi_{B}$. Then, the difference in photo-electrons $n_e$ can be determined by signal subtraction through the measuring of current. 
Coherent detection scheme employed in this work is similar to homodyne detection. Homodyning in SCW-CV is carried out directly in the phase modulator in the Bob module (instead o a $50/50$ beam splitter) for each of the sidebands independently. After the second modulation interference is observed at frequencies $\omega_{k}=\omega+k\Omega$ if equal microwave field frequencies $\Omega$ are used by Alice and Bob. Resulting carrier and subcarriers wave power depends on phase difference between $\varphi_{A}$ and $\varphi_{B}$. In case of constructive (Fig. \ref{fig1.1}a) or destructive (Fig. \ref{fig1.1}b) interference, subcarriers wave power becomes either more or less than the carrier wave power, respectively. A narrow spectral filter then separates the carrier from the sidebands. Finally the two output modes (carrier and all the sidebands) are detected by two different photodiodes, and their photo currents are subtracted. Thus, one can extract information encoded in the the phase of the oscillating signal. Similar to traditional homodyne detection in QKD, Bob measures only one quadrature component at a time.

\subsection*{\label{sec:part22}Subcarrier wave CV-QKD protocol}

The protocol consists of the following steps:
\begin{enumerate}
    \item Alice prepares a multimode coherent state $|\psi_0(\varphi_A)\rangle=\bigotimes_{k=-S}^S|{\alpha_k(\varphi_A)}\rangle_k$ by choosing from a finite set of states (4 states is in our case). She assumes $|\psi_0(0)\rangle$, $|\psi_0(\pi/2)\rangle$ as "0" and $|\psi_0(\pi)\rangle$, $|\psi_0(3\pi/2)\rangle$ as "1".
    \item Bob measures the received state in one of two bases: $x_k$ or $p_k$. We define the quadratures as:
    \begin{equation}
    \begin{aligned}
        x_k=\frac{1}{2}(a_k+a^{\dagger}_k)\\p_k=\frac{1}{2i}(a_k-a^{\dagger}_k),
    \end{aligned}
    \end{equation}
    where  $a_k$ and $a_k^{\dagger}$ are annihilation and creation operators for mode $k$, respectively, with commutation relation $[a_{\nu},a_{\mu}^{\dagger}]=\delta_{\nu,\mu}$. Commutator for $x_k$ and $p_k$  is expressed as $[x_k,p_k]=i/2$, so the uncertainty relation is $\delta x_k\delta p_k \geq 1/4$.

  Interference in the Bob's module (see "Quantum state preparation" and "Detection" sections) occurs between the corresponding modes, thus the described relations are not violated. Bob measures one of the quadratures by homodyne detection methods. The procedures described above are repeated required (large) number of times.
    \item For each time instance, Alice and Bob reveal their selected bases, and mismatched bases are discarded. Bob forms his bit string by assigning 0 for negative $v$ and 1 for the positive  $v$ values in measurement results. The threshold values are selected to maximize the secure key rate.
    \item Alice and Bob apply error correction and privacy amplification procedures. In this paper, we consider only the case of direct reconciliation (DR), when Bob adjusts his data in accordance with the data of Alice. As a result, the secure secret key is distributed.
\end{enumerate}

\subsection*{\label{sec:part23}Quantum state preparation}

The states prepared by Alice can be described in terms of representation basis of abelian cyclic point symmetry groups $C_M$ respectively. The protocol which we propose here is based on four coherent states (number of bases $N=2$). The initial state at Alice's side is $\ket{\sqrt{\mu_{0}}}_0 \otimes \ket{vac}_{SB}$, where $\ket{vac}_{SB}$ is the sidebands vacuum state and $\ket{\sqrt{\mu_{0}}}_0$ is the carrier wave coherent state with the average number of photons $\mu_{0}$ emitted from a coherent monochromatic light source with  frequency $\omega$.

The state at the Alice's modulator output is a multimode coherent state

\begin{equation}
|\psi_0(\varphi_A)\rangle = \bigotimes_{k=-S}^S|{\alpha_k(\varphi_A)}\rangle_k,
\end{equation}
with coherent amplitudes
\begin{equation}\label{alpha}
\alpha_k(\varphi_A)=\sqrt{\mu_0}d^S_{0k}(\beta_A)e^{-i(\theta_1+\varphi_A)k},
\end{equation}
where $\theta_1$ is a constant phase and $d^S_{nk}(\beta_A)$ is the Wigner d-function that appears in the quantum theory of angular momentum \cite{Varshalovich1988}. Argument of the d-function $\beta_A$ is determined by the Alice's modulation index $m_A$, disregarding the modulator medium dispersion this dependence can be written as
\begin{equation}\label{beta}
\cos{({\beta_A})}=1-\frac{1}{2}{\left(\frac{m_A}{S+0.5}\right)^2}.
\end{equation}
The detailed description of electro-optic modulation process for quantum states can be found in \cite{Miroshnichenko2017}.
\subsection*{\label{sec:part24}Detection}
The traveling wave phase modulator on the Bob's side has the same modulation frequency $\Omega$
as in the Alice's one, but a different phase $\varphi_B$ and modulation index $m_B$. The resulting state is also a multimode coherent state
\begin{equation}\label{psiB}
|\psi_B(\varphi_A,\varphi_B)\rangle = \bigotimes_{k=-S}^S|{\alpha_k'(\varphi_A,\varphi_B)}\rangle_k,
\end{equation}
with coherent amplitudes
\begin{equation}\label{alphaprime}
\alpha_k'(\varphi_A,\varphi_B)=\sqrt{\mu_0\eta(L)}\exp(-i\theta_2k) d^S_{0k}(\beta'),
\end{equation}
where $\eta(L)$ is the transmission coefficient of the quantum channel. New argument of the d-function is
\begin{equation}\label{betaprime}
\cos{\beta'}=\cos{\beta_A}\cos{\beta_B}-\sin{\beta_A}\sin{\beta_B}\cdot\cos\left(\varphi_A-\varphi_B+\varphi_0\right),
\end{equation}
where $\theta_2$ and $\varphi_0$ are phases determined by phase modulator structure \cite{Miroshnichenko2017}. In order to achieve constructive interference, Bob should use $\varphi_0$ as an offset for his phase and apply microwave phase $\varphi=\varphi_0+\varphi_B$ in his modulator.
According to \cite{Kozubov2017} the average number of photons arriving at the first arm of Bob's detector in the transmission window $T$ is
\begin{equation}
\label{eq:n1}
n_{1}\left(\varphi_{A}, \varphi_{B}\right)=\mu_{0} \eta(L) \eta_{B}\left(1-(1-\vartheta)\left|d_{00}^{S}\left(\beta^{\prime}\right)\right|^{2}\right),
\end{equation}
$\eta_{B}$ is the losses in Bob's module and $\vartheta$ is carrier wave attenuation factor. Thus the average number of photons arriving at the second arm of Bob's detector is
\begin{equation}
\label{eq:n2}
n_{2}\left(\varphi_{A}, \varphi_{B}\right)=\mu_{0} \eta(L) \eta_{B}(1-\vartheta)\left|d_{00}^{S}\left(\beta^{\prime}\right)\right|^{2},
\end{equation}
After simple mathematical manipulations, we obtain
\begin{equation}
\beta^{\prime}=\beta_A\sqrt{\left(\delta^2+2\delta\cos \left(\varphi_{A}-\varphi_{B}+\varphi_0\right)+1\right)},
\end{equation}
where $\delta=\beta_B/\beta_A$.

Then, depending on Bob's phase choice $\varphi_{B}$, the photon number difference is proportional to photo-electron number difference and therefore to the measured quadrature value. The normalized quadrature value of the signal is obtained as
\begin{equation}
    v = \frac{(n_1\left(\varphi_{A}, \varphi_{B}\right)-n_2\left(\varphi_{A}, \varphi_{B}\right))\cdot s}{2 \cdot\sqrt{n_{LO}}},
\end{equation}
where $n_{LO}$ is mean number of photons on the carrier before the second phase modulation, and $s$ is detector sensitivity.

\begin{figure}[ht!]
   \centering
     \includegraphics[width=0.5\textwidth]{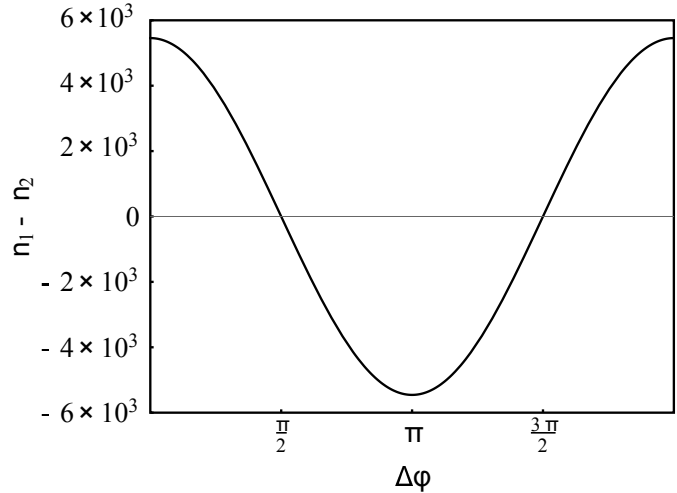}
 \caption{Dependence of the mean photon number difference on the relative phase shift represented by a cosine function. In this case the difference is maximal at points $0$ and $\pi$ and equals zero at points $\pi / 2$ and $3\pi/2$.}
 \label{ncos}
 \end{figure}

When bases coincide the power arriving at Bob's detectors will be greater either at its first or second arm, depending on the phase difference. The argument of d-function $\beta_A$ (and, subsequently, modulation index) is determined by mean photon number which is selected to maximise secure key rate. Parameter $\delta$, as a ratio of modulation indices, is optimized in order to achieve the same distinguishability of quadratures for different phases in a correctly chosen basis, so that \mbox{$(n_{1}\left(0, 0\right)-n_{2}\left(0, 0\right))=|n_{1}\left(\pi, 0\right)-n_{2}\left(\pi, 0\right)|$}. Hence, Bob observes quadrature distributions that are symmetrically offset with respect to zero. The dependence of mean number of photons on the relative phase shift is illustrated in Figure~\ref{ncos}.
\subsection*{\label{sec:part25}Quantum bit error rate}
\begin{figure}[ht!]
   \centering
     \includegraphics[width=0.5\textwidth]{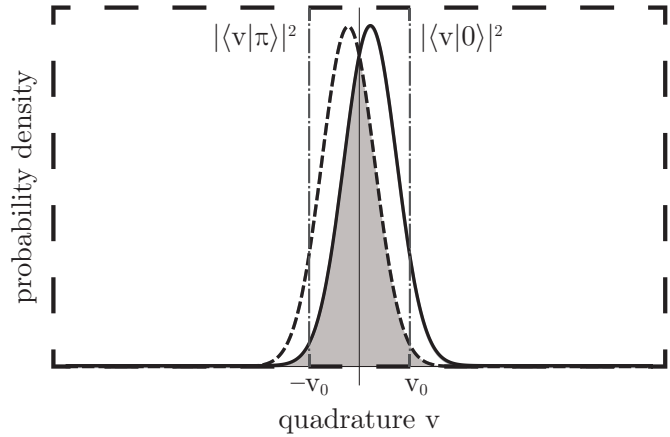}
 \caption{Quadrature distributions for correct basis with threshold values $\{-v_0,v_0\}$ with $\varphi_{A}=\varphi_{B}=\pi$.}
 \label{fig:dist}
 \end{figure}

Succeeding the detection stage for pulses in correct bases we obtain two probability density distributions (Fig.~\ref{fig:dist}) that contain information about binary signals. Our channel is characterised by excess noise variance $\Xi$ and vacuum noise variance, which is constantly defined as $V=1/4$  \cite{Symul2007, Heid}. So, the probability density to obtain $v$ conditioned by the amplitude $\alpha_k$ is:

\begin{equation}
\displaystyle
p(v|\varphi_A + \varphi)=\sqrt{\frac{2}{\pi(1+\Xi)}}e^{-2{\frac{(v-\sqrt{\eta(L)}\alpha_k)^2}{1+\Xi}}},
\end{equation}

The overlap between the distributions contributes to the bit errors. Bob can set the threshold value $v_0$ in order to reduce the number of errors, then Bob expects "0", if $v<-v_0$ and "1", if $v>v_0$, thereby increasing inconclusive result. Therefore for each choice of basis it has two input values, Alice's bits $x=\{0, 1\}$, and three output values: Bob's bits $y=\{0, 1\}$ and an inconclusive result or $y=?$. 
Considering the quantum channel as a binary symmetric channel~(BSC), one may estimate detection probability density~$(1-g)$, where $(g)$ is erasure, and the probability density that Bob assigns the wrong bit value~$(e)$, in other words, if $\varphi_{A}=\varphi_{B}=\pi$ we obtain:
\begin{equation}
1-g=p(0|\varphi)+p(0|\pi + \varphi),
\end{equation}
\begin{equation}
e=\frac{p (0|\pi + \varphi)}{p(0| \varphi)+p(0|\pi + \varphi)}.
\end{equation}
After the post-selection stage, we can calculate bit error rate as $Q=E/P$, where the error probability $E$ and post-selection rate $P$, respectively, are obtained as follows
\begin{equation}
E=\int_{-\infty}^{v_0} e(v) dv,
\end{equation}
\begin{equation}
P=\int_{-v_0}^{v_0} (1-g(v)) dv.
\end{equation}

\subsection*{\label{sec:part33}Holevo bound}
Let us consider a collective attack in the asymptotic limit on infinitely long keys for the case of our system and compute the corresponding asymptotic collective key rate using the Devetak-Winter approach \cite{Devetak2005}. We estimate an upper bound for Eve's knowledge about the data using Holevo bound \cite{Holevo1973} for weak coherent states. Finite-key analysis for our protocol is presented in the following section.

Here we use direct reconciliation scheme. In this case Alice sends error correction information to Bob and the secret key is determined by Alice's data. Eve can rotate all states stored in her quantum memory after reconciliation and before her measurement. Holevo bound can be found considering unconditioned channel density operator. The Eve's quantum state, conditioned on Alice's data, is
\begin{equation}\label{psi}
|\psi_E(\varphi_A)\rangle=|\psi_0(\varphi_A)\rangle. 
\end{equation}
Eve needs to discriminate between the states in one basis
\begin{equation}\label{rho}
\rho=\frac{1}{2}|\psi_E(0)\rangle\langle\psi_E(0)|+\frac{1}{2}|\psi_E(\pi)\rangle\langle\psi_E(\pi)|.
\end{equation}
The Holevo bound is given by
\begin{equation}
\chi_{DR}=S(\rho)-\sum_{j} p_jS(\rho_j),
\end{equation}
where $S(\rho)$ is the von Neumann entropy, index j enumerates the possible states in the quantum channel, $\rho_j$ is the ancilla state under condition that $j$th state was attacked, $p_j$ is the weight of the $j$th state. The von Neumann entropy of a density operator is the Shannon entropy of its eigenvalues. The eigenvalues of the channel density operator $\rho$ are
\begin{equation} \label{lambda}
\lambda_{1,2} =\frac{1}{2}\left(1\pm |\langle\psi(0)|\psi(\pi)\rangle|\right).
\end{equation}
The overlapping of our states can be described as
\begin{equation}
\langle\psi(0)|\psi(\pi)\rangle=\exp \left[-\mu_{0}\left(1-d^S_{00}(2\beta_A)\right)\right].
\end{equation}
We therefore obtain the Holevo bound using binary Shannon entropy function $h(x)$:
\begin{equation}\label{chi2}
\begin{aligned} 
\chi_{DR} = h\left(\frac12(1-\exp \left[-\mu_{0}\left(1-d^S_{00}(2\beta_A)\right)\right]\right).
\end{aligned} 
\end{equation}

\begin{figure}[t]
   \centering
     \includegraphics[width=0.6\textwidth]{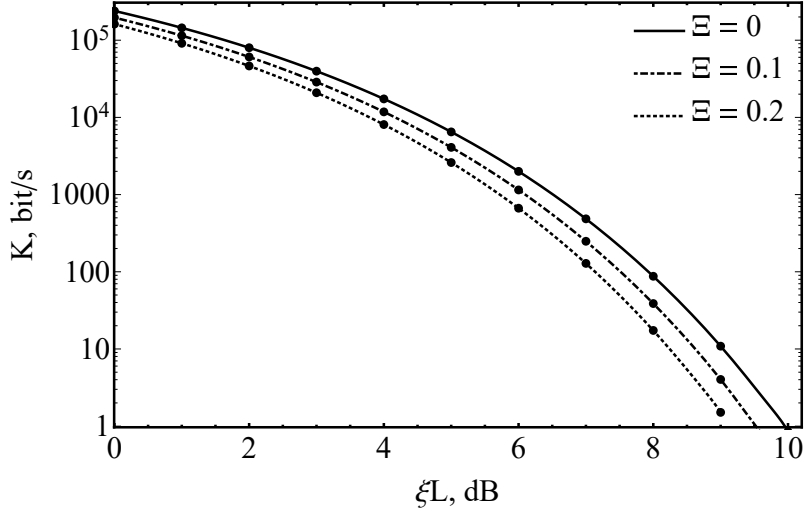}
 \caption{Secure key rate $K$ dependence on channel loss in SCW CV-QKD system with discrete modulation including two cases of asymptotic key: with excess noise $\Xi=0$, $\Xi=0.1$ and $\Xi=0.2$.}
 \label{K}
 \end{figure}

Now we are able to estimate the secure key generation rate~$K$:
\begin{equation}
\label{oldkeyrate}
K=\int_{v_0}^{\infty}\frac{(1-g)}{NT}\left[1-h\left(e\right)-\chi \right]dv.
\end{equation}
The secret key rates as functions of channel loss are shown in Fig.~\ref{K}. The parameters of the system are \mbox{$T=100$ ns}, \mbox{$\eta_{B}=10^{-0.64}$}, \mbox{$\theta=10^{-6}$}, \mbox{$\varphi_0=5^{\circ}$}. We consider the ideal case and the case of the excess noise variance $\Xi=0.1$. The parameters $\mu$, $\mu_0$ and $v_0$ are optimized so as to maximize the secret key rate. The value $v_0$ was optimized for losses at various distances.
 Equation~\eqref{oldkeyrate} describes only the asymptotic case of infinitely long key sequences. In order to evaluate real keys it makes sense to carry out another estimation taking into account finite-key effects.
\subsection*{\label{sec:part35}Secure key generation rate with finite-key effects}
\begin{figure}[ht]
   \centering
     \includegraphics[width=0.6\textwidth]{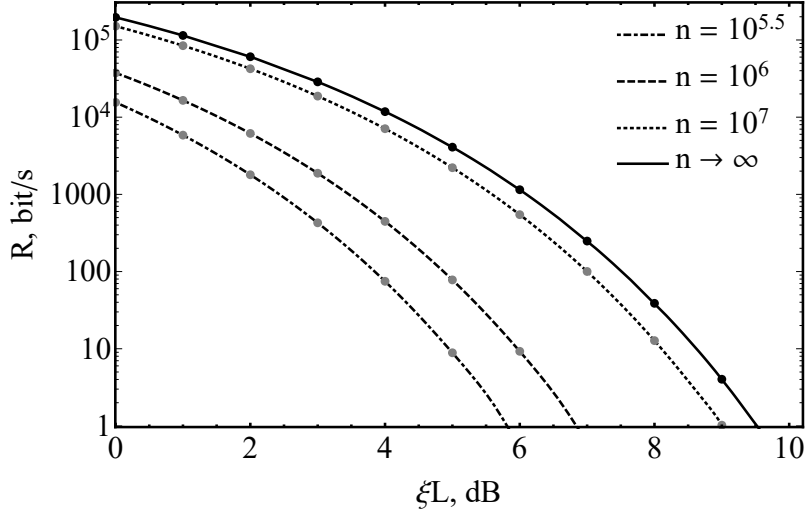}
 \caption{Secure key rate $R$ dependence on channel loss in SCW CV-QKD system with discrete modulation for different number of detected quantum bits $n$.}
 \label{R}
 \end{figure}
To estimate appropriate bound on secure key rate we consider the notation of Rényi entropies since they describe the worst case and not the average one \cite{renyi1961, Renner2008}. We bound $\epsilon$-smooth min-entropy \cite{renyi1961, Tomamichel2012, Kozubov2019} as follows:
\begin{equation} \label{eq:hminnew}
H^{\varepsilon_s}_{min}(\textbf{A}|\textbf{E})\geq n\left(H(\textbf{A}|\textbf{E})-\frac{\delta(\varepsilon_S)}{\sqrt{n}}\right),
\end{equation}
where 
\begin{equation} \label{eq:deltas}
\delta(\varepsilon_s)=4\log{(2+\sqrt{2})}\sqrt{\log{\left(\frac{2}{\varepsilon^2_s}\right)}},
\end{equation}
here and $H(\textbf{A}|\textbf{E})$ is conditional von Neumann entropy and it denotes the entropy of Alice’s bit conditioned on Eve’s side-information in a
single round, Eve's side information is \textbf{E}. Conditional von Neumann entropy in case of direct reconciliation can be bounded as \mbox{$H(\textbf{A}|\textbf{E})\ge 1-\chi_{DR}$}.
On the error correction step both parties should check and remover the errors in their bit strings. Here we assume that Alice and Bob use \mbox{low-density parity-check} (LDPC) codes \cite{Gallager1962}. Bob randomly chooses a $k$ bits and sends them to Alice, then Alice estimates the quantum channel parameters. It should be noted that LDPC codes succeed only if the actual error rate value $Q_{real}$ is less than a reference value parameterized in the code. Thus, Alice needs to consider an additional error rate fraction \mbox{$\Delta Q$}. It can be estimated in order to maximize the probability of successful error correction in one round while keeping the secret key rate as high as possible. Then Alice computes the syndrome of LDPC code that corrects up to \mbox{$n(Q_{est}+\Delta Q)$} error bits. We denote the length of the syndrome as 
\begin{equation} \label{codeapprox}
code_{EC}\approx nf_{EC}h(Q_{est}+\Delta Q),
\end{equation}
where $f_{EC}$ is error correction efficiency. Using the syndrome, Bob corrects the bits forming some new bit string $\textbf{B'}$ and applies a two-universal hash function with output length $check_{EC}$. Bob then sends the hash to Alice in order to check whether their strings match. If the hashes are different, Alice enlarges $\Delta Q$ or aborts the protocol. Otherwise Alice obtains the bit string $\textbf{A'}$. The remaining smooth-entropy is 
\begin{equation} \label{eq:hmin}
H^{\varepsilon_s}_{min}(\textbf{A'}|\textbf{E})\geq n\left(H(\textbf{A}|\textbf{E})-\frac{\delta(\varepsilon_S)}{\sqrt{n}}\right)-k-code_{EC}-check_{EC},
\end{equation}
where sample size $k$ is estimated by maximizing the key rate \cite{Kozubov2019}. At privacy amplification step Alice and Bob hash their bit strings to a key of length $l$ \cite{Arnon2016,Kozubov2019}
\begin{equation} \label{eq:l}
l=n\left(H(\textbf{A}|\textbf{E})-\frac{\delta(\varepsilon_S)}{\sqrt{n}}\right)-k-code_{EC}-check_{EC}-loss_{PA},
\end{equation}
At the error correction step, we have to estimate "correctness error" $\varepsilon_{EC}$. From the properties of 2-universal hashing $\varepsilon_{EC}$ is
\begin{equation} \label{eq:eps}
\varepsilon_{EC}=2^{-check_{EC}},
\end{equation}
The trace distance $d$ between the protocol output and an ideal output is bounded by $d\leq\varepsilon_{s}+\varepsilon_{PA}$. We therefore obtain that the protocol is \mbox{$\varepsilon_{QKD}$-secure and correct} protocol, with \mbox{$\varepsilon_{QKD}=\varepsilon_{EC}+\varepsilon_{s}+\varepsilon_{PA}$}.
Finally, the dependence of average secret key rates on losses in the quantum channel for different values of n is
\begin{equation}
\label{newkeyrate}
\begin{aligned} R &=\int_{v}^{\infty}\frac{1-g}{NT} \cdot \bigg(1-\chi-4 \frac{1}{\sqrt{n}} \log (2+\sqrt{2}) \sqrt{\log \left(\frac{2}{\varepsilon_{s}^{2}}\right)}\\ &-\frac{1}{n}\left(k+\operatorname{code}_{EC}+\log \frac{1}{\varepsilon_{E C}}+\log \frac{1}{\varepsilon_{P A}}-2\right) \bigg)dv. \end{aligned}
\end{equation}

It should be noted that in the asymptotic case $n\to\infty$, the equations \eqref{oldkeyrate} and \eqref{newkeyrate} converge to the same expression.
The secret key rates for different values of $n$ are presented in Fig.~\ref{R} as a function of channel loss. The parameters $\mu$, $\mu_0$, $k$ and $v_0$ are optimized so as to maximize the secret key rate. The value $v_0$ is also optimized for losses at various distances. The considered security parameters are as follows: $\varepsilon_s=\varepsilon_{PA}=10^{-10}$, $\varepsilon_{EC}=2^{-256}$.
\section*{\label{sec:part4}Discussion}
In this paper we proposed the implementation of CV-QKD protocol using SCW method, build a mathematical model of the proposed scheme and demonstrate the security proof technique. We calculated the secure key rate for discrete modulation \mbox{CV-QKD} protocol with post-selection in the asymptotic and finite-size regime. We calculated the lower bound on the secret key rate for the \mbox{CV-QKD} system under the assumption that the quantum channel noise is negligible compared to detector noise and Eve is restricted to collective attacks. Our calculation shows that the system allows to provide a secret key for channel losses up to 9 dB in a realistic system implementation. It is important to note that our scheme also allows to implement \mbox{CV-QKD} with Gaussian modulation and the presented security analysis can be adopted there. 
Subsequent works will focus on a full security proof, as well as the experimental implementation of the proposed protocol. 
\section*{\label{sec:part5}Acknowledgements}
This work was funded by Government of Russian Federation (grant MK-777.2020.8).
\end{document}